\documentclass[final,3p,times]{elsarticle}

\usepackage{amssymb}
\usepackage{amsmath}
\usepackage{xspace}
\usepackage{microtype}
\usepackage{placeins}

\newcommand{\pp}           {pp\xspace}

\newcommand{\PbPb}         {\mbox{Pb--Pb}\xspace}
\newcommand{\pA}           {\mbox{p--A}\xspace}
\newcommand{\pPb}          {\mbox{p--Pb}\xspace}

\newcommand{\snn}          {\ensuremath{\sqrt{s_{\mathrm{NN}}}}\xspace}
\newcommand{\pt}           {\ensuremath{p_{\rm T}}\xspace}
\newcommand{\meanpt}       {$\langle p_{\mathrm{T}}\rangle$\xspace}

\newcommand{\etarange}[1]  {\mbox{$\left | \eta \right |<#1$}}
\newcommand{\yrange}[1]    {\mbox{$\left | y \right |<#1$}}

\newcommand{\dndeta}       {\ensuremath{\mathrm{d}N_\mathrm{ch}/\mathrm{d}\eta}\xspace}

\newcommand{\acceff}       {\ensuremath{A\kern-.15em\times\kern-.15em\varepsilon}\xspace}

\newcommand{\nineH}        {$\sqrt{s}~=~0.9$~Te\kern-.1emV\xspace}
\newcommand{\seven}        {$\sqrt{s}~=~7$~Te\kern-.1emV\xspace}
\newcommand{\twoH}         {$\sqrt{s}~=~0.2$~Te\kern-.1emV\xspace}
\newcommand{\twosevensix}  {$\sqrt{s}~=~2.76$~Te\kern-.1emV\xspace}
\newcommand{\five}         {$\sqrt{s}~=~5.02$~Te\kern-.1emV\xspace}
\newcommand{\twosevensixnn}{$\sqrt{s_{\mathrm{NN}}}~=~2.76$~Te\kern-.1emV\xspace}
\newcommand{\fivenn}       {$\sqrt{s_{\mathrm{NN}}}~=~5.02$~Te\kern-.1emV\xspace}

\newcommand{\GeVc}         {Ge\kern-.1emV/$c$\xspace}
\newcommand{\MeVc}         {Me\kern-.1emV/$c$\xspace}
\newcommand{\TeV}          {Te\kern-.1emV\xspace}
\newcommand{\GeV}          {Ge\kern-.1emV\xspace}
\newcommand{\MeV}          {Me\kern-.1emV\xspace}
\newcommand{\GeVmass}      {Ge\kern-.2emV/$c^2$\xspace}
\newcommand{\MeVmass}      {Me\kern-.2emV/$c^2$\xspace}

\newcommand{\kzero}        {\ensuremath{{\rm K}^{0}_{\rm{S}}}\xspace}
\newcommand{\lmb}          {\ensuremath{\Lambda}\xspace}
\newcommand{\almb}         {\ensuremath{\overline{\Lambda}}\xspace}
\newcommand{\Om}           {\ensuremath{\Omega^-}\xspace}
\newcommand{\Mo}           {\ensuremath{\overline{\Omega}^+}\xspace}
\newcommand{\X}            {\ensuremath{\Xi^-}\xspace}
\newcommand{\Ix}           {\ensuremath{\overline{\Xi}^+}\xspace}

\newcommand{\XiNosign}           {\ensuremath{\Xi}\xspace}
\newcommand{\OmNosign}           {\ensuremath{\Omega}\xspace}

\newcommand{\avdndetaplot} {\ensuremath{{\langle\dndeta\rangle}_{\etarange{0.5}}}\xspace}

\biboptions{sort&compress}
\journal{Journal of Subatomic Particles and Cosmology}

\begin{document}

\begin{frontmatter}

\title{Strangeness production in light-ion collisions with ALICE at the LHC}

\author[]{Sara \textsc{Pucillo}\textsuperscript{a} for the ALICE Collaboration}

\affiliation[aaa]{
  organization={Università degli Studi di Salerno, and Sezione INFN Salerno},
  addressline={Via Giovanni Paolo II, 132},
  city={Fisciano (SA)},
  postcode={84084},
  country={Italy}
}

\begin{abstract}
Measurements in pp and \pA collisions have revealed that small collision systems exhibit most of the signs traditionally attributed to heavy-ion collisions, such as the smooth increase of the strange hadron yields with the collision multiplicity (strangeness enhancement). A key question is how these effects evolve with system size and whether they can be described within a unified framework.

The recently collected oxygen--oxygen (OO) collision data by ALICE at \snn = 5.36~\TeV provide an unprecedented opportunity to explore an intermediate-size system that naturally bridges the gap between pp and \PbPb collisions. In this contribution, we present the first results on the production of (multi-)strange particles as a function of charged-particle multiplicity in OO collisions. This allows us to investigate strangeness enhancement across different system sizes at comparable multiplicities, providing new insights into the mechanisms of strangeness production. The experimental results are compared with model predictions from various Monte Carlo generators.
\end{abstract}


\end{frontmatter}


\section{Introduction} \label{sec1}

The enhanced production of strange hadrons in heavy-ion collisions with respect to minimum-bias \pp collisions has historically been interpreted as one of the signatures of the formation of a deconfined quark--gluon plasma, known as \textit{strangeness enhancement}~\cite{rafelski}. A major result of the LHC heavy-ion program is the observation by the ALICE Collaboration that the strange-hadron-to-pion yield ratios increase smoothly with charged-particle multiplicity, from low- to high-multiplicity \pp and \pPb collisions, reaching values close to those measured in peripheral \PbPb collisions~\cite{nature,pp5,pp7,pp13,pPbV0,pPbCasc,PbPb2,PbPb5}. This behaviour suggests that final-state multiplicity plays a central role in strangeness production. However, the microscopic origin of the enhancement in small collision systems remains unclear, and current QCD-inspired Monte Carlo generators do not provide a quantitative description of the observed trends. Additional experimental constraints are therefore needed to disentangle the role of multiplicity from possible effects related to the collision geometry, system size, and hadronization dynamics.

In this context, OO collisions provide a particularly relevant system. They bridge the gap between small and large collision systems, reaching charged-particle multiplicities comparable to those observed in high-multiplicity \pp and \pPb events, while having a different initial geometry and collision volume. The study of (multi-)strange particle production in OO collisions therefore offers an ideal opportunity to test whether the smooth evolution observed from \pp to \PbPb reflects a common underlying mechanism, or whether deviations emerge when the system size and initial geometry are changed at similar multiplicity. 

\section{Strange hadron production in OO collisions} \label{sec2}

In ALICE~\cite{Acharya_2024}, strange and multi-strange hadrons are reconstructed at midrapidity, \yrange 0.5, through their characteristic weak-decay topologies. The \kzero and \lmb candidates are reconstructed from V0 decays, while \XiNosign and \OmNosign baryons are reconstructed through cascade decays; particles and antiparticles are analysed separately, but charge signs are omitted in the following notation when referring to both states. Charged-particle tracking and particle identification are performed with the central barrel detectors, mainly the Inner Tracking System (ITS) and the Time Projection Chamber (TPC). The selected tracks are combined to build invariant-mass distributions, and topological and kinematic selections are applied to suppress the combinatorial background~\cite{topological}.

The raw yields are extracted from fits to the invariant-mass distributions. The signal peak is described with a Gaussian function for \kzero and \lmb, and with a double-sided Crystal Ball function for \XiNosign and \OmNosign. The background is modelled with particle-dependent functions: an exponential for \kzero, a first-order polynomial for \lmb, and a second-order polynomial for \XiNosign and \OmNosign. The signal is counted in the interval $\mu \pm 5\sigma$ for strange hadrons and $\mu \pm 8\sigma$ for multi-strange hadrons, after subtracting the fitted background contribution.

The raw yields are corrected for acceptance, reconstruction and selection efficiencies, signal losses induced by the event selection, event-normalisation effects, and possible duplicate reconstructed events:
\begin{equation}
\frac{1}{N_{\mathrm{evt}}^{\mathrm{corr}}}
\frac{d^2N_{\mathrm{corr}}}{d\pt d\textit{y}}
=
\frac{1}{N_{\mathrm{evt}}/\epsilon_{\mathrm{normalization}}}
\frac{d^2N_{\mathrm{raw}}}{d\pt d\textit{y}}
\frac{1}{\epsilon_{\mathrm{signal}}},
\label{eq:corrections1}
\end{equation}
where $\epsilon_{\mathrm{normalization}}$ accounts for the event-normalisation correction and $\epsilon_{\mathrm{signal}}$ includes the signal-related efficiency corrections.

The analysis is performed as a function of event activity, defined using percentiles of the signal amplitude measured by the Fast Interaction Trigger (FIT) detectors. The FT0 detector consists of two arrays of Cherenkov detectors covering forward pseudorapidity regions: FT0A at $3.5 < \eta < 4.9$ and FT0C at $-3.3 < \eta < -2.1$. In this analysis, the FT0C amplitude is used as the event-activity estimator, and the corresponding classes are referred to as FT0C centrality percentiles. Each percentile class is associated with the corresponding average charged-particle multiplicity density at midrapidity, \avdndetaplot.

\subsection{Experimental results} \label{subsec2}

\begin{figure}[!htbp]
  \centering
  \includegraphics[width=0.48\linewidth]{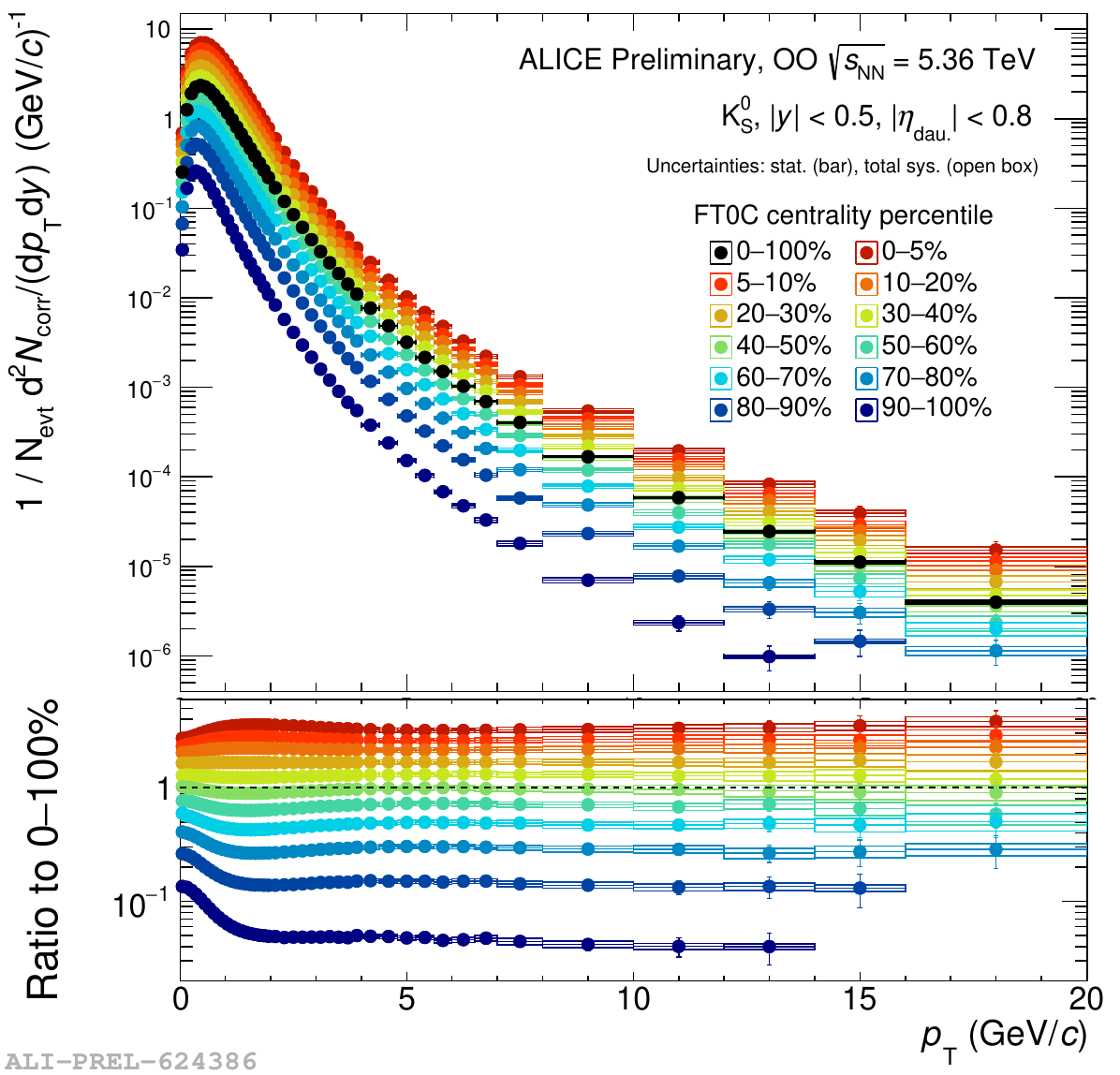}
  \hfill
  \includegraphics[width=0.48\linewidth]{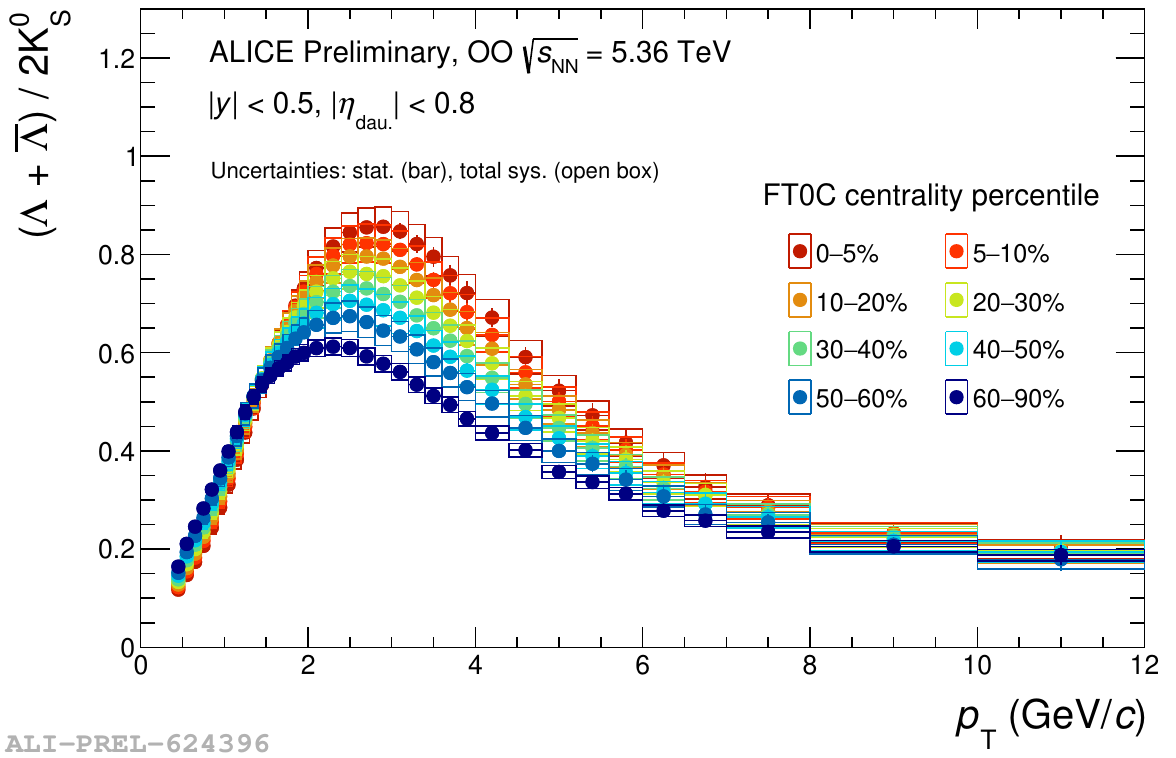}
  \caption{(\textit{left}) Corrected \pt-differential spectra of \kzero in OO collisions at \snn = 5.36~\TeV for different centrality classes. The lower panel shows the ratio to the centrality-integrated 0--100\% result. (\textit{right}) Corrected $(\lmb+\almb)/(2\kzero)$ ratio as a function of \pt in OO collisions at \snn = 5.36~\TeV for different centrality classes. In both panels, statistical uncertainties are represented by vertical bars, while total systematic uncertainties are shown as open boxes.}
  \label{Fig1}
\end{figure}

Figure~\ref{Fig1} shows, on the left, the first centrality-differential measurement of corrected \kzero \pt-differential spectra in OO collisions at \snn = 5.36~\TeV. The spectra exhibit a clear evolution with centrality: moving from peripheral to central collisions, the distributions become harder, with a less steep fall-off and an enhanced relative contribution at intermediate and high \pt. This trend is consistent with stronger radial collective effects in higher-multiplicity events. The lower panel reports the ratios of the centrality-differential spectra to the centrality-integrated 0--100\% result. These ratios highlight sizeable variations among centrality classes, especially at low \pt, while a flatter behaviour is observed above approximately 3~\GeVc. Similar results have also been obtained for \lmb+\almb, \X+\Ix and \Om+\Mo hadrons.

To further investigate possible radial-flow effects in OO collisions, the $(\lmb+\almb)/(2\kzero)$ ratio has been studied as a function of \pt for different centrality classes, as shown in Fig.~\ref{Fig1} on the right. The ratio exhibits a pronounced intermediate-\pt enhancement, with a centrality-dependent maximum in the range \pt $\simeq$ 2--3~\GeVc; the enhancement becomes more prominent and shifts towards larger \pt when moving from peripheral to central collisions. This centrality evolution is qualitatively consistent with the presence of collective radial expansion, which affects heavier baryons more strongly and pushes their spectral yield towards higher transverse momentum as the event multiplicity increases. In this \pt region, the observed baryon-to-meson enhancement~\cite{PbPb2} may reflect the interplay between collective expansion and hadronization mechanisms, such as quark recombination or coalescence~\cite{
MinissaleCoalescence}. At high \pt, the ratios measured in the different centrality classes tend to converge, indicating a reduced sensitivity to centrality and a possible transition towards a regime where fragmentation-like processes play a dominant role.

\begin{figure}[!htbp]
  \centering
  \includegraphics[width=0.48\linewidth]{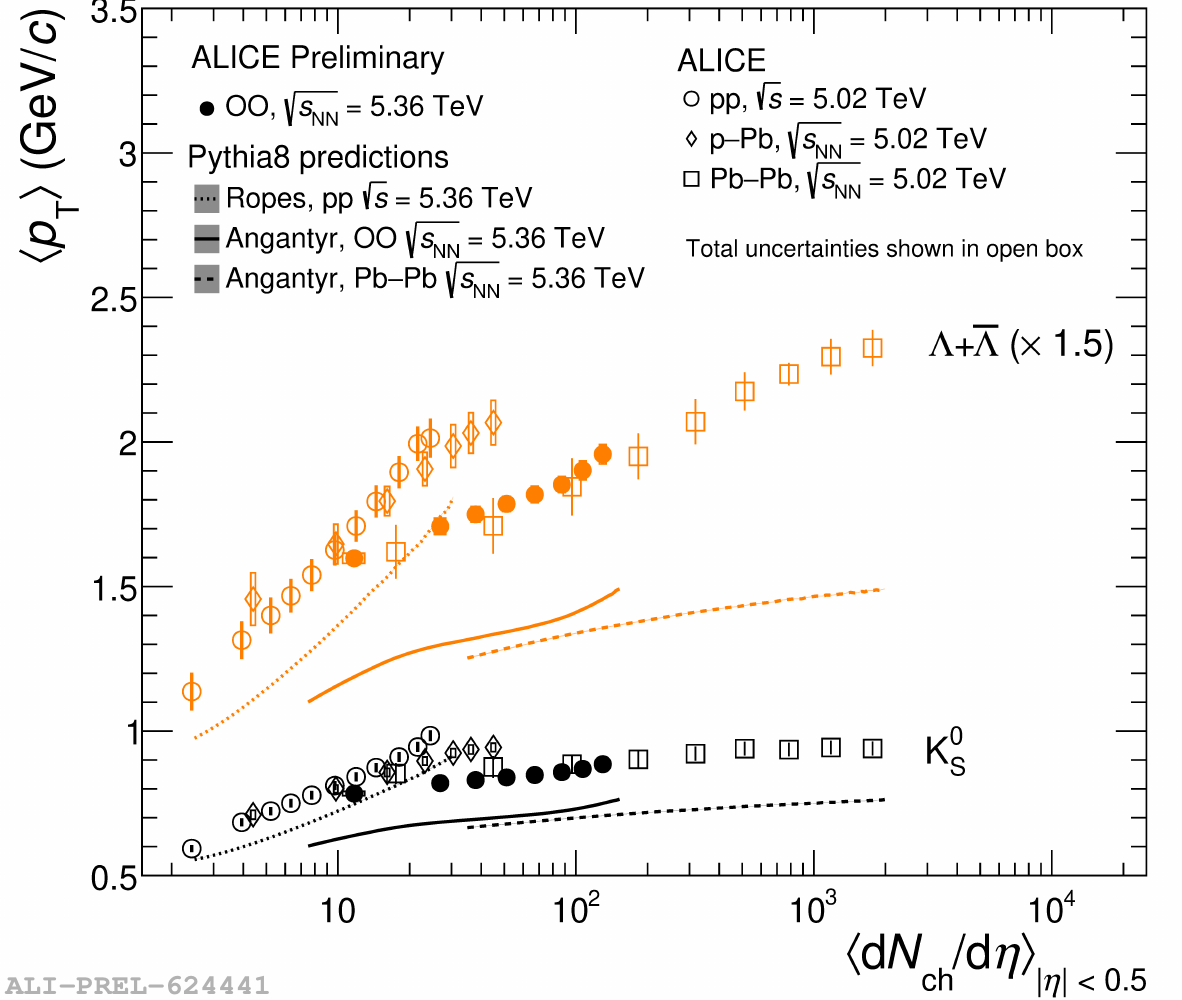}
  \hfill
  \includegraphics[width=0.48\linewidth]{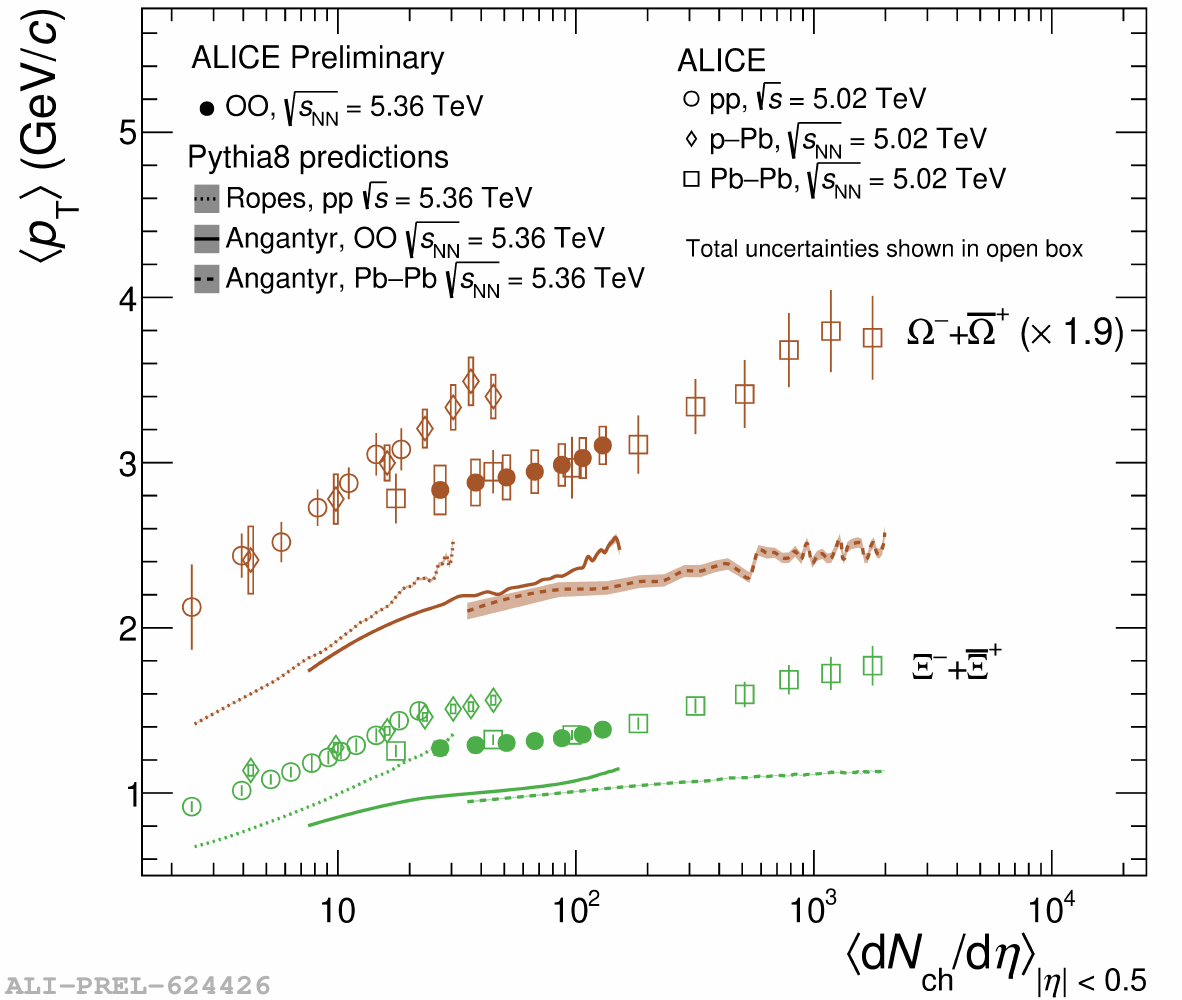}
  \caption{Mean transverse momentum as a function of the charged-particle multiplicity for strange (left) and multi-strange (right) hadrons in OO collisions (solid markers) at \snn = 5.36~\TeV, compared with \pp~\cite{pp5}, \pPb~\cite{pPbV0,pPbCasc}, and \PbPb~\cite{PbPb5} results at \snn = 5.02~\TeV (open markers). Open boxes represent total uncertainties. Model calculations from Pythia8 Angantyr~\cite{Angantyr} and Ropes~\cite{Ropes} are also shown.}
  \label{Fig2}
\end{figure}

To quantify the evolution of the spectral shape with multiplicity, the mean transverse momentum, \meanpt, has been extracted in OO collisions as a function of the charged-particle multiplicity density \avdndetaplot. The results are shown in Fig.~\ref{Fig2}: the left panel reports the measurements for strange particles, \kzero in black and \lmb+\almb in orange, while the right panel shows the corresponding results for multi-strange particles, \X+\Ix in green and \Om+\Mo in brown. The OO results at \snn = 5.36~\TeV, shown as solid markers, are compared with measurements in \pp~\cite{pp5}, \pPb~\cite{pPbV0,pPbCasc}, and \PbPb~\cite{PbPb5} collisions at \snn = 5.02~\TeV, shown as open markers.
For all particle species, \meanpt increases with multiplicity, reflecting the progressive hardening of the \pt spectra. The increase is stronger for heavier hadrons, consistently with the mass-dependent behaviour expected from radial collective expansion. Although the OO measurements cover a multiplicity range comparable to small collision systems, their evolution follows more closely the \PbPb trend, showing that similar final-state multiplicities do not necessarily imply similar particle-production dynamics. 
The results are further compared with available model predictions, namely Pythia8 Angantyr~\cite{Angantyr} for OO and \PbPb collisions and Pythia8 Ropes~\cite{Ropes} for \pp collisions. These models do not fully reproduce both the magnitude and the multiplicity dependence of the measured \meanpt values, providing additional constraints on the modelling of spectral hardening across different collision systems.

\begin{figure}[!htbp]
  \centering
  \includegraphics[width=0.42\linewidth]{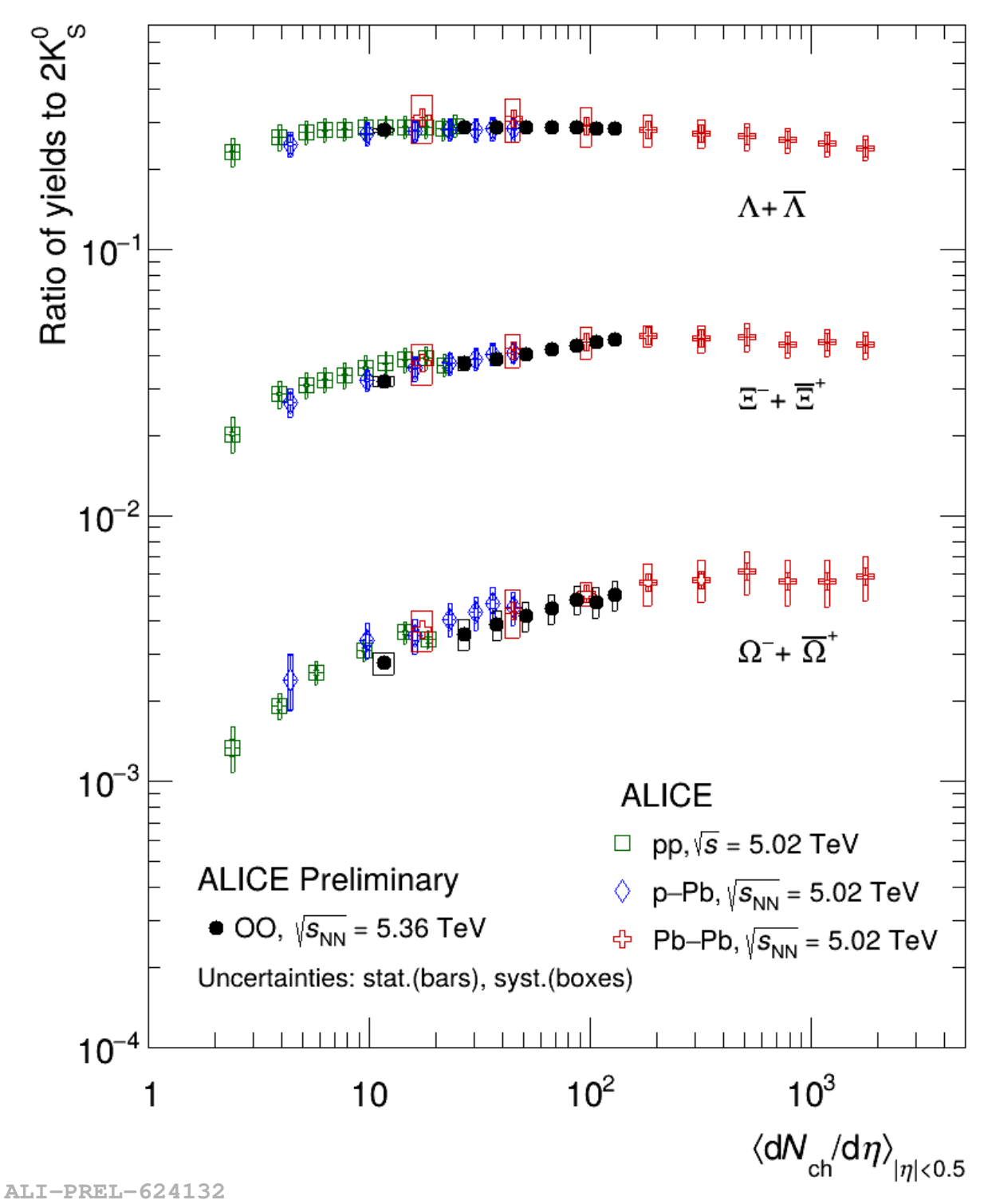}
  \caption{Ratios of \pt-integrated \lmb+\almb, \X+\Ix, and \Om+\Mo yields to 2\kzero as a function of the charged-particle multiplicity density in OO collisions at \snn = 5.36~\TeV (black solid markers), compared with \pp~\cite{pp5}, \pPb~\cite{pPbV0,pPbCasc}, and \PbPb~\cite{PbPb5} results at \snn = 5.02~\TeV. Statistical and systematic uncertainties are shown as vertical bars and boxes, respectively.}
  \label{Fig3}
\end{figure}

Finally, Figure~\ref{Fig3} shows the ratios of \lmb+\almb, \X+\Ix, and \Om+\Mo yields to 2\kzero as a function of the charged-particle multiplicity density in different collision systems. The OO results follow the smooth multiplicity-dependent evolution observed in \pp~\cite{pp5}, \pPb~\cite{pPbV0,pPbCasc}, and \PbPb~\cite{PbPb5} collisions, confirming that charged-particle multiplicity is an effective scaling variable for relative strange-hadron production. Together with the centrality-dependent spectral hardening, the baryon-to-meson ratios, and the \meanpt evolution, these results show that OO collisions provide a bridge between small and large collision systems. While integrated yield ratios are largely governed by multiplicity, the dynamical observables indicate that similar final-state multiplicities do not necessarily correspond to the same hadronization dynamics.

\FloatBarrier
\bibliographystyle{elsarticle-num}
\bibliography{sqm2026_template}

@article{rafelski,
    author = "Rafelski, Johann and M{\"u}ller, Berndt",
    title = "",
    journal = "Phys. Rev. Lett.",
    volume = "56",
    pages = "2334",
    year = "1986"
}

@article{PbPb2,
    author = "{B. Abelev, et al. [ALICE Collaboration]}",
    title = "",
    archivePrefix = "arXiv",
    eprint = "1307.5543",
    primaryClass = "nucl-ex",
    journal = "Phys. Lett. B",
    volume = "728",
    pages = "216--227",
    year = "2014"
}

@article{nature,
    author = "{J. Adam, et al. [ALICE Collaboration]}",
    title = "",
    archivePrefix = "arXiv",
    eprint = "1606.07424",
    primaryClass = "nucl-ex",
    journal = "Nature Phys.",
    volume = "13",
    pages = "535--539",
    year = "2017"
}

@article{pp7,
    author = "{S. Acharya, et al. [ALICE Collaboration]}",
    title = "",
    archivePrefix = "arXiv",
    eprint = "1807.11321",
    primaryClass = "nucl-ex",
    journal = "Phys. Rev. C",
    volume = "99",
    pages = "024906",
    year = "2019"
}

@article{pp13,
    author = "{S. Acharya, et al. [ALICE Collaboration]}",
    title = "",
    archivePrefix = "arXiv",
    eprint = "1908.01861",
    primaryClass = "nucl-ex",
    journal = "Eur. Phys. J. C",
    volume = "80",
    pages = "167",
    year = "2020"
}

@article{pPbV0,
    author = "{B. Abelev, et al. [ALICE Collaboration]}",
    title = "",
    archivePrefix = "arXiv",
    eprint = "1307.6796",
    primaryClass = "nucl-ex",
    journal = "Phys. Lett. B",
    volume = "728",
    pages = "25--38",
    year = "2014"
}

@article{pPbCasc,
    author = "{J. Adam, et al. [ALICE Collaboration]}",
    title = "",
    archivePrefix = "arXiv",
    eprint = "1512.07227",
    primaryClass = "nucl-ex",
    journal = "Phys. Lett. B",
    volume = "758",
    pages = "389--401",
    year = "2016"
}

@article{PbPb5,
    author = "{I. J. Abualrob, et al. [ALICE Collaboration]}",
    title = "",
    archivePrefix = "arXiv",
    eprint = "2511.10360",
    primaryClass = "nucl-ex",
    year = "2025"
}

@article{pp5,
    author = "{I. J. Abualrob, et al. [ALICE Collaboration]}",
    title = "",
    archivePrefix = "arXiv",
    eprint = "2511.10306",
    primaryClass = "nucl-ex",
    year = "2025"
}

@article{Acharya_2024,
    author = "{S. Acharya, et al. [ALICE Collaboration]}",
    title = "",
    archivePrefix = "arXiv",
    eprint = "2302.01238",
    journal = "JINST",
    volume = "19",
    pages = "P05062",
    year = "2024"
}

@article{topological,
    author = "{K. Aamodt, et al. [ALICE Collaboration]}",
    title = "",
    archivePrefix = "arXiv",
    eprint = "1012.3257",
    primaryClass = "hep-ex",
    journal = "Eur. Phys. J. C",
    volume = "71",
    pages = "1594",
    year = "2011"
}

@article{Angantyr,
    author = "Bierlich, Christian and Gustafson, G{\"o}sta and L{\"o}nnblad, Leif and Shah, Harsh",
    title = "",
    archivePrefix = "arXiv",
    eprint = "1806.10820",
    primaryClass = "hep-ph",
    journal = "JHEP",
    volume = "10",
    pages = "134",
    year = "2018"
}

@article{Ropes,
    author = "Bierlich, Christian and Gustafson, G{\"o}sta and L{\"o}nnblad, Leif and Tarasov, Andrey",
    title = "",
    archivePrefix = "arXiv",
    eprint = "1412.6259",
    primaryClass = "hep-ph",
    journal = "JHEP",
    volume = "03",
    pages = "148",
    year = "2015"
}

@article{MinissaleCoalescence,
    author        = "Minissale, Vincenzo and Scardina, Francesco and Greco, Vincenzo",
    title         = "",
    journal       = "Phys. Rev. C",
    volume        = "92",
    pages         = "054904",
    year          = "2015",
    archivePrefix = "arXiv",
    eprint        = "1502.06213",
    primaryClass  = "nucl-th"
}

\end{document}